# Every Sequence is Decompressible from a Random One

David Doty[*]
Department of Computer Science
Iowa State University
Ames, IA 50011 USA
`ddoty` *at* `iastate` *dot* `edu`

**Abstract**

Kučera and Gács independently showed that every infinite sequence is Turing reducible to a Martin-Löf random sequence. This result is extended by showing that every infinite sequence $S$ is Turing reducible to a Martin-Löf random sequence $R$ such that the asymptotic number of bits of $R$ needed to compute $n$ bits of $S$, divided by $n$, is precisely the constructive dimension of $S$. It is shown that this is the optimal ratio of query bits to computed bits achievable with Turing reductions. As an application of this result, a new characterization of constructive dimension is given in terms of Turing reduction compression ratios.

## 1 Introduction

An (infinite, binary) sequence $S$ is Turing reducible to a sequence $R$, written $S \leq_\mathrm{T} R$, if there is an algorithm $M$ that can compute $S$, given oracle access to $R$. Any computable sequence is trivially Turing reducible to any other sequence. Thus, if $S \leq_\mathrm{T} R$, then intuitively we can consider $R$ to contain the uncomputable information that $M$ needs to compute $S$.

Informally, a sequence is Martin-Löf random [Mar66] if it has no structure that can be detected by any algorithm. Kučera [Kuč85, Kuč89] and Gács [Gác86] independently obtained the surprising result that *every* sequence is Turing reducible to a Martin-Löf random sequence. Thus, it is possible to store information about an arbitrary sequence

[*]This research was funded in part by grant number 9972653 from the National Science Foundation as part of their Integrative Graduate Education and Research Traineeship (IGERT) program.

$S$ into another sequence $R$, while ensuring that the storage of this information imparts no detectable structure on $R$. In the words of Gács, "it permits us to view even very pathological sequences as the result of the combination of two relatively well-understood processes: the completely chaotic outcome of coin-tossing, and a transducer algorithm." Merkle and Mihailović [MM04] have provided a simpler proof of this result using martingales, which are strategies for gambling on successive bits of a sequence.

Bennett [Ben88] claims that "This is the infinite analog of the far more obvious fact that every finite string is computable from an algorithmically random string (e.g., its minimal program)." However, the analogy is incomplete. Not only is every string $s$ computable from a random string $r$, but $r$ is an *optimally compact representation* of $s$. Viewing the sequence $R$ as a compressed representation of the sequence $S$, the asymptotic number of bits of $R$ needed to compute $n$ bits of $S$, divided by $n$, defines the compression ratio between them. Gács showed that his reduction achieves a compression ratio of 1: for any $n$, $n + o(n)$ bits of $R$ are required to compute $n$ bits of $S$. But as in the case of strings, sequences that are sparse in information content should in principle be derivable from a more compact description.

Lutz [Lut03b] defined the *(constructive) dimension* $\dim(S)$ of a sequence $S$ as an effective version of Hausdorff dimension (the most widely-used fractal dimension; see [Hau19, Fal90]). Constructive dimension is a measure of the "density of computably enumerable information" in a sequence. Lutz defined dimension in terms of constructive *gales*, a generalization of martingales. Mayordomo [May02] proved that for all sequences $S$, $\dim(S) = \liminf_{n \to \infty} \frac{K(S \restriction n)}{n}$, where $K(S \restriction n)$ is the Kolmogorov complexity of the $n^{\text{th}}$ prefix of $S$.

Athreya et al. [AHLM06], also using gales, defined the *(constructive) strong dimension* $\Dim(S)$ of a sequence $S$ as an effective version of packing dimension (see [Tri82, Sul84, Fal90]), another type of fractal dimension and a dual of Hausdorff dimension. They proved the analogous characterization $\Dim(S) = \limsup_{n \to \infty} \frac{K(S \restriction n)}{n}$. Since Kolmogorov complexity is a lower bound on the algorithmic compression of a finite string, $\dim(S)$ and $\Dim(S)$ can respectively be considered to measure the best- and worst-case compression ratios achievable on finite prefixes of $S$.

Consider the following example. It is well known that $K$, the characteristic sequence of the halting language, has dimension and strong dimension 0 [Bar68]. The binary representation of Chaitin's halting probability $\Omega = \sum_{M \text{ halts}} 2^{-|M|}$ (where $M$ ranges over all halting programs and $|M|$ is $M$'s description length) is an algorithmically random sequence [Cha75]. It is known that $K \leq_{\text{T}} \Omega$ (see [LV97]). Furthermore, only the first $n$ bits of $\Omega$ are required to compute the first $2^n$ bits of $K$, so the asymptotic compression ratio of this reduction is 0. $\Omega$ can be considered an optimally compressed representation of $K$, and it is no coincidence that the compression ratio of 0 achieved by the reduction is precisely the dimension of $K$.

We generalize this phenomenon to arbitrary sequences, extending the result of Kučera and Gács by pushing the compression ratio of the reduction down to its optimal lower bound. Thus, this paper completes Bennett's above-mentioned analogy between reductions to random sequences and reductions to random strings. Compression can be mea-sured by considering both the best- and worst-case limits of compression, corresponding respectively to measuring the limit inferior and the limit superior of the compression ratio on longer and longer prefixes of $S$. We show that, for every sequence $S$, there is a sequence $R$ such that $S \leq_T R$, where the best-case compression ratio of the reduction is the dimension of $S$, and the worst-case compression ratio is the strong dimension of $S$. Furthermore, we show that the sequence $R$ can be chosen to be Martin-Löf random, although the randomness of $R$ is easily obtained by invoking the construction of Gács in a black-box fashion. The condition that $R$ is random is introduced chiefly to show that our main result is a strictly stronger statement than the result of Kučera and Gács, but the compression is the primary result. Finally, a single machine works in all cases; as is the case with Kolmogorov complexity, a single Turing reduction reproduces each sequence $S$ from its shortest description. Our result also extends a compression result of Ryabko [Rya84, Rya86], discussed in section 3, although it is not a strict improvement, since Ryabko considered two-way reductions (Turing equivalence) rather than one-way reductions.

One application of this result is a new characterization of constructive dimension as the optimal compression ratio achievable on a sequence with Turing reductions. This compression characterization differs from Mayordomo's Kolmogorov complexity characterization in that the compressed version of a prefix of $S$ does not change drastically from one prefix to the next, as it would in the case of Kolmogorov complexity. While the theory of Kolmogorov complexity assigns to each finite string an optimally compact representation of that string – its shortest program – this does not easily allow us to compactly represent an infinite sequence with another infinite sequence. This contrasts, for example, the notions of finite-state compression [Huf59] or Lempel-Ziv compression [ZL78], which are *monotonic*: for all strings $x$ and $y$, $x \sqsubseteq y$ ($x$ is a prefix of $y$) implies that $C(x) \sqsubseteq C(y)$, where $C(x)$ is the compressed version of $x$. Monotonicity enables these compression algorithms to encode and decode an infinite sequence – or in the real world, a data stream of unknown length – online, without needing to reach the end of the data before starting. However, if we let $\pi(x)$ and $\pi(y)$ respectively be shortest programs for $x$ and $y$, then $x \sqsubseteq y$ does not imply that $\pi(x) \sqsubseteq \pi(y)$. In fact, it may be the case that $\pi(x)$ is longer than $\pi(y)$, or that $\pi(x)$ and $\pi(y)$ do not even share any prefixes in common. In the self-delimiting formulation of Kolmogorov complexity, $\pi(x)$ *cannot* be a prefix of $\pi(y)$.

Our characterization of sequence compression via Turing reductions, coupled with the fact that the optimal compression ratio is always achievable by a single oracle sequence and reduction machine, gives a way to associate with each sequence $S$ another sequence $R$ that is an optimally compressed representation of $S$. As in the case of Kolmogorov complexity, the compression direction is in general uncomputable; it is not always the case that $R \leq_T S$.

# 2 Preliminaries

## 2.1 Notation

All logarithms are base 2. We write $\mathbb{R}$, $\mathbb{Q}$, $\mathbb{Z}$, and $\mathbb{N}$ for the set of all reals, rationals, integers, and non-negative integers, respectively. For $A \subseteq \mathbb{R}$, $A^+$ denotes $A \cap (0, \infty)$.

$\{0,1\}^*$ is the set of all finite, binary *strings*. The length of a string $x \in \{0,1\}^*$ is denoted by $|x|$. $\lambda$ denotes the empty string. Let $\sigma_0, \sigma_1, \sigma_2, \ldots \in \{0,1\}^*$ denote the standard enumeration of binary strings $\sigma_0 = \lambda, \sigma_1 = 0, \sigma_2 = 1, \sigma_3 = 00, \ldots$. For $k \in \mathbb{N}$, $\{0,1\}^k$ denotes the set of all strings $x \in \{0,1\}^*$ such that $|x| = k$. The Cantor space $\mathbf{C} = \{0,1\}^\infty$ is the set of all infinite, binary *sequences*. For $x \in \{0,1\}^*$ and $y \in \{0,1\}^* \cup \mathbf{C}$, $xy$ denotes the concatenation of $x$ and $y$, and $x \sqsubseteq y$ denotes that $x$ is a *prefix* of $y$; i.e., there exists $u \in \{0,1\}^* \cup \mathbf{C}$ such that $xu = y$. For $S \in \{0,1\}^* \cup \mathbf{C}$ and $i, j \in \mathbb{N}$, we write $S[i]$ to denote the $i^{\text{th}}$ bit of $S$, with $S[0]$ being the leftmost bit, we write $S[i \mathinner{\ldotp\ldotp} j]$ to denote the substring consisting of the $i^{\text{th}}$ through $j^{\text{th}}$ bits of $S$ (inclusive), with $S[i \mathinner{\ldotp\ldotp} j] = \lambda$ if $i > j$, and we write $S \upharpoonright i$ to denote $S[0 \mathinner{\ldotp\ldotp} i-1]$.

## 2.2 Kolmogorov Complexity

We work with the self-delimiting Kolmogorov complexity. See [LV97] for an account of this model. All Turing machines are *self-delimiting*. This means that

- a Turing machine $M$ is allowed to move its input tape head only to the right, and

- if $M$ does not halt with its tape head on the rightmost bit of its input, the computation is considered invalid.

Fix a self-delimiting universal Turing machine $U$. Let $x \in \{0,1\}^*$. The *Kolmogorov complexity* of $x$ is
$$\mathrm{K}(x) = \min_{\pi \in \{0,1\}^*} \{\, |\pi| \mid U(\pi) = x \,\}.$$

For all $q \in \mathbb{Q}$, let $\mathrm{K}(q) = \mathrm{K}(s(q))$, where $s(q) \in \{0,1\}^*$ is some standard binary representation of the rational $q$ with a numerator, denominator, and sign bit.

For all $w \in \{0,1\}^*$, let $e_0(w) = 0^{|w|}1w$. Define the *self-delimiting encoding function* $\mathrm{enc} : \{0,1\}^* \to \{0,1\}^*$ for all $w \in \{0,1\}^*$ by

$$\mathrm{enc}(w) = e_0\left(\sigma_{|w|}\right) w.$$

For all $n \in \mathbb{N}$, let $\mathrm{enc}(n) = \mathrm{enc}(\sigma_n)$.

Strings encoded by enc and valid programs for $U$ are *self-delimiting*. They can be prepended to arbitrary strings and uniquely decoded.

*Observation* 2.1. For all $x \in \{0,1\}^*$, $|\mathrm{enc}(x)| \leq |x| + 2\log|x| + 3$, and for all $n \in \mathbb{N}$, $\mathrm{enc}(n) \leq \log n + 2 \log \log n + 3$.

Our results, being asymptotic in nature, do not depend crucially on using the self-delimiting Kolmogorov complexity K; it is simply more convenient for encoding purposes. All results would work out the same if we instead use the plain Kolmogorov complexity $C : \{0,1\}^* \to \mathbb{N}$ (see [LV97]). Whenever we would need to add a program to a string and retain the ability to uniquely decode it, we could simply encode the program using the function enc.

## 2.3 Reductions and Compression

Let $M$ be a Turing machine and $S \in \mathbf{C}$. We say $M$ *computes* $S$ if, on input $n \in \mathbb{N}$, $M$ outputs the string $S \upharpoonright n$.

We define an *oracle Turing machine* ($OTM$) to be a Turing machine $M$ that can make constant-time queries to an oracle sequence, and we let OTM denote the set of all oracle Turing machines. For $R \in \mathbf{C}$, we say $M$ operates *with oracle $R$* if, whenever $M$ makes a query to index $n \in \mathbb{N}$, the bit $R[n]$ is returned.

Let $S, R \in \mathbf{C}$ and $M \in \text{OTM}$. We say $S$ *is Turing reducible to $R$ via $M$*, and we write $S \leq_T R$ via $M$, if $M$ computes $S$ with oracle $R$. In this case, define $M(R) = S$. We say $S$ *is Turing reducible to $R$*, and we write $S \leq_T R$, if there exists $M \in \text{OTM}$ such that $S \leq_T R$ via $M$.

Since we do not consider space or time bounds with Turing reductions, we may assume without loss of generality that an oracle Turing machine queries each bit of the oracle sequence at most once, caching the bit for potential future queries.

Let $S, P, R \in \mathbf{C}$ and $M_S^P, M_P^R \in \text{OTM}$ such that $S \leq_T P$ via $M_S^P$ and $P \leq_T R$ via $M_P^R$. Define the *composition of $M_S^P$ with $M_P^R$*, denoted $M_S^P \circ M_P^R$, to be the oracle Turing machine that works as follows. On input $n \in \mathbb{N}$ and with oracle $R$, $M_S^P \circ M_P^R$ simulates $M_S^P$ to compute $S \upharpoonright n$. Whenever a bit of $P$ is queried by $M_S^P$, $M_S^P \circ M_P^R$ simulates $M_P^R$ with oracle $R$ for the minimum number of steps needed to compute that bit of $P$.

*Observation* 2.2. $\leq_T$ is transitive: if $S \leq_T P$ via $M_S^P$ and $P \leq_T R$ via $M_P^R$, then $S \leq_T R$ via $M_S^P \circ M_P^R$.

In order to view Turing reductions as decompression algorithms, we must define how to measure the amount of compression achieved. Let $S, R \in \mathbf{C}$ and $M \in \text{OTM}$ such that $S \leq_T R$ via $M$. Define $\#_S^R(M, n)$ to be the *query usage of $M$ on $S \upharpoonright n$ with oracle $R$*, the number of bits of $R$ queried by $M$ when computing $S \upharpoonright n$.[1] Define

$$\rho_M^-(S, R) = \liminf_{n \to \infty} \frac{\#_S^R(M, n)}{n},$$

$$\rho_M^+(S, R) = \limsup_{n \to \infty} \frac{\#_S^R(M, n)}{n}.$$

$\rho_M^-(S, R)$ and $\rho_M^+(S, R)$ are respectively the best- and worst-case compression ratios as $M$ decompresses $R$ into $S$. Note that $0 \leq \rho_M^-(S, R) \leq \rho_M^+(S, R) \leq \infty$. Let $S \in \mathbf{C}$. The *lower*

---
[1]If we instead defined $\#_S^R(M, n)$ to be the index of the rightmost queried bit (i.e., assuming that if a bit is queried, all bits to the left of it are also queried), all results of the present paper would still hold.

and upper compression ratios of $S$ are respectively defined

$$\rho^-(S) = \min_{\substack{R \in \mathbf{C} \\ M \in \text{OTM}}} \left\{ \rho_M^-(S, R) \mid S \leq_\text{T} R \text{ via } M \right\},$$

$$\rho^+(S) = \min_{\substack{R \in \mathbf{C} \\ M \in \text{OTM}}} \left\{ \rho_M^+(S, R) \mid S \leq_\text{T} R \text{ via } M \right\}.$$

Note that $0 \leq \rho^-(S) \leq \rho^+(S) \leq 1$. As we will see, by Lemma 4.2 and Theorem 4.3, the two minima above exist. In fact, there is a single OTM $M$ that achieves the minimum compression ratio in each case.

## 2.4 Constructive Dimension

See [Lut03a, Lut03b, AHLM06, Lut05] for a more comprehensive account of the theory of constructive dimension and other effective dimensions.

1. An *s-gale* is a function $d : \{0, 1\}^* \to [0, \infty)$ such that, for all $w \in \{0, 1\}^*$,

$$d(w) = 2^{-s}[d(w0) + d(w1)].$$

2. A *martingale* is a 1-gale.

Intuitively, a martingale is a strategy for gambling in the following game. The gambler starts with some initial amount of *capital* (money) $d(\lambda)$, and it reads an infinite sequence $S$ of bits. $d(w)$ represents the capital the gambler has after reading the prefix $w \sqsubseteq S$. Based on $w$, the gambler bets some fraction of its capital that the next bit will be 0 and the remainder of its capital that the next bit will be 1. The capital bet on the bit that appears next is doubled, and the remaining capital is lost. The condition $d(w) = \frac{d(w0)+d(w1)}{2}$ ensures *fairness*: the martingale's expected capital after seeing the next bit, given that it has already seen the string $w$, is equal to its current capital. The fairness condition and an easy induction lead to the following observation.

*Observation* 2.3. Let $k \in \mathbb{N}$ and let $d : \{0, 1\}^* \to [0, \infty)$ be a martingale. Then

$$\sum_{u \in \{0,1\}^k} d(u) = 2^k d(\lambda).$$

An $s$-gale is a martingale in which the capital bet on the bit that occurred is multiplied by $2^s$, as opposed to simply 2, after each bit. The parameter $s$ may be regarded as the *unfairness of the betting environment*; the lower the value of $s$, the faster money is taken away from the gambler. Let $d : \{0, 1\}^* \to [0, \infty)$ be a martingale and let $s \in [0, \infty)$. Define the *s-gale induced by $d$*, denoted $d^{(s)}$, for all $w \in \{0, 1\}^*$ by

$$d^{(s)}(w) = 2^{(s-1)|w|} d(w).$$

If a gambler's martingale is given by $d$, then, for all $s \in [0, \infty)$, its $s$-gale is $d^{(s)}$.

The following theorem, due to Lutz, establishes an upper bound on the number of strings on which an $s$-gale can perform well.

**Theorem 2.4.** [Lut03a] *Let $d$ be an $s$-gale. Then for all $w \in \{0,1\}^*$, $k \in \mathbb{N}$, and $\alpha \in \mathbb{R}^+$, there are fewer than $\frac{2^k}{\alpha}$ strings $u \in \{0,1\}^k$ for which*

$$\max_{v \sqsubseteq u} \left\{ 2^{(1-s)|v|} d(wv) \right\} \geq \alpha d(w).$$

**Corollary 2.5.** *Let $d$ be a martingale. Then for all $l \in \mathbb{R}$, $w \in \{0,1\}^*$, $k \in \mathbb{N}$, and $\alpha \in \mathbb{R}^+$, there are fewer than $\frac{2^l}{\alpha}$ strings $u \in \{0,1\}^k$ for which*

$$d(wu) \geq \alpha 2^{k-l} d(w).$$

Let $S \in \mathbf{C}$, $s \in [0, \infty)$, and let $d : \{0,1\}^* \to [0, \infty)$ be an $s$-gale. $d$ *succeeds on $S$*, and we write $S \in \mathrm{S}^\infty[d]$, if

$$\limsup_{n \to \infty} d(S \upharpoonright n) = \infty.$$

$d$ *strongly succeeds on $S$*, and we write $S \in \mathrm{S}^\infty_{\mathrm{str}}[d]$, if

$$\liminf_{n \to \infty} d(S \upharpoonright n) = \infty.$$

An $s$-gale succeeds on $S$ if, for every amount of capital $C \in \mathbb{R}^+$, it eventually makes capital at least $C$. An $s$-gale strongly succeeds on $S$ if, for every amount of capital $C$, it eventually makes capital at least $C$ *and* stays above $C$ forever.

Let $d : \{0,1\}^* \to [0, \infty)$ be an $s$-gale. We say that $d$ is *constructive (a.k.a. lower semicomputable, subcomputable)* if there is a computable function $\widehat{d} : \{0,1\}^* \times \mathbb{N} \to \mathbb{Q}$ such that, for all $w \in \{0,1\}^*$ and $t \in \mathbb{N}$,

1. $\widehat{d}(w, t) \leq \widehat{d}(w, t+1) < d(w)$, and

2. $\lim_{t \to \infty} \widehat{d}(w, t) = d(w)$.

Let $R \in \mathbf{C}$. We say that $R$ is *Martin-Löf random*, and we write $R \in \mathrm{RAND}$, if there is no constructive martingale $d$ such that $R \in \mathrm{S}^\infty[d]$. This definition of Martin-Löf randomness, due to Schnorr [Sch71], is equivalent to Martin-Löf's traditional definition (see [Mar66, LV97]).

The following well-known theorem (see [MM04]) says that there is a *single* constructive martingale that *strongly* succeeds on every $S \notin \mathrm{RAND}$.

**Theorem 2.6.** [MM04] *There is a constructive martingale $\mathbf{d}$ such that $\mathrm{S}^\infty_{\mathrm{str}}[\mathbf{d}] = \mathrm{RAND}^c$.*

Let $\widehat{\mathbf{d}} : \{0,1\}^* \times \mathbb{N} \to \mathbb{Q}$ be the computable function testifying that $\mathbf{d}$ is constructive.

The following theorem, due independently to Hitchcock and Fenner, states that $\mathbf{d}^{(s)}$ is "optimal" for the class of constructive $t$-gales whenever $s > t$.

**Theorem 2.7.** [Hit03, Fen02] *Let $s > t \in \mathbb{R}^+$, and let $d$ be a constructive $t$-gale. Then*

$$\mathrm{S}^\infty[d] \subseteq \mathrm{S}^\infty[\mathbf{d}^{(s)}] \text{ and } \mathrm{S}^\infty_{\mathrm{str}}[d] \subseteq \mathrm{S}^\infty_{\mathrm{str}}[\mathbf{d}^{(s)}].$$

By Theorem 2.7, the following definition of constructive dimension is equivalent to the definitions given in [Lut03b, AHLM06]. Let $X \subseteq \mathbf{C}$. The *constructive dimension* and the *constructive strong dimension* of $X$ are respectively defined

$$\begin{aligned} \mathrm{cdim}(X) &= \inf\{s \in [0, \infty) \mid X \subseteq S^\infty[\mathbf{d}^{(s)}]\}, \\ \mathrm{cDim}(X) &= \inf\{s \in [0, \infty) \mid X \subseteq S^\infty_{\mathrm{str}}[\mathbf{d}^{(s)}]\}. \end{aligned}$$

Let $S \in \mathbf{C}$. The *dimension* and the *strong dimension* of $S$ are respectively defined

$$\begin{aligned} \dim(S) &= \mathrm{cdim}(\{S\}), \\ \mathrm{Dim}(S) &= \mathrm{cDim}(\{S\}). \end{aligned}$$

Intuitively, the (strong) dimension of $S$ is the *most unfair betting environment $s$* in which the optimal constructive gambler $\mathbf{d}$ (strongly) succeeds on $S$.

*Observation* 2.8. Let $S \in \mathbf{C}$. If $s > \dim(S)$ and $s' > \mathrm{Dim}(S)$, then for infinitely many $n$, $\mathbf{d}(S \upharpoonright n) \geq 2^{(1-s)n}\mathbf{d}(\lambda)$, and for all but finitely many $n$, $\mathbf{d}(S \upharpoonright n) \geq 2^{(1-s')n}\mathbf{d}(\lambda)$.

*Observation* 2.9. If $S \in \mathrm{RAND}$, then $\dim(S) = \mathrm{Dim}(S) = 1$.

The following theorem – the first part due to Mayordomo [May02] and the second to Athreya et al. [AHLM06] – gives a useful characterization of the dimension of a sequence in terms of Kolmogorov complexity, and it justifies the intuition that dimension measures the *density of computably enumerable information* in a sequence.

**Theorem 2.10.** [May02, AHLM06] *For all $S \in \mathbf{C}$,*

$$\dim(S) = \liminf_{n \to \infty} \frac{\mathrm{K}(S \upharpoonright n)}{n}, \text{ and } \mathrm{Dim}(S) = \limsup_{n \to \infty} \frac{\mathrm{K}(S \upharpoonright n)}{n}.$$

One of the most important properties of constructive dimension is that of *absolute stability*, shown by Lutz [Lut03b], which allows us to reason equivalently about the constructive dimension of individual sequences and sets of sequences:

**Theorem 2.11.** [Lut03b] *For all $X \subseteq \mathbf{C}$,*

$$\mathrm{cdim}(X) = \sup_{S \in X} \dim(S), \text{ and } \mathrm{cDim}(X) = \sup_{S \in X} \mathrm{Dim}(S).$$

## 3 Previous Work

The next theorem says that every sequence is Turing reducible to a random sequence. Part 1 is due independently to Kučera and Gács, and part 2 is due to Gács.

**Theorem 3.1.** [Kuč85, Kuč89, Gác86] *There is an OTM $M$ such that, for all $S \in \mathbf{C}$, there is a sequence $R \in \mathrm{RAND}$ such that*

1. $S \leq_{\mathrm{T}} R$ via $M$.

2. $\rho^+_M(S, R) = 1$.

Let $X \subseteq \mathbf{C}$. Define the *code cost* of $X$ by

$$c_{\mathrm{T}}(X) = \inf_{M_e, M_d \in \mathrm{OTM}} \left\{ \sup_{S \in X} \rho^-_{M_d}(S, M_e(S)) \;\middle|\; (\forall S \in X)\, M_d(M_e(S)) = S \right\}.$$

$c_\mathrm{T}(X)$ is the optimal lower compression ratio achievable with *reversible* Turing reductions on sequences in $X$. The next theorem is due to Ryabko [Rya84, Rya86].

**Theorem 3.2.** [Rya86] *For every $X \subseteq \mathbf{C}$, $c_\mathrm{T}(X) = \mathrm{cdim}(X)$.*

Ryabko defined $c_\mathrm{T}$ based on what he calls "$T$-codes" and did not explicitly mention OTMs, but these are essentially equivalent. A $T$-code is a pair of encoder/decoder (i.e. compressor/decompressor) algorithms $E, D : \{0,1\}^* \to \{0,1\}^*$ – implemented by the Turing machines $M_e$ and $M_d$ in the present paper's definition of $c_\mathrm{T}$ – which are required to be *monotonic*: for all $x, y \in \{0,1\}^*$,

$$x \sqsubseteq y \implies E(x) \sqsubseteq E(y) \text{ and } D(x) \sqsubseteq D(y).$$

$M_e$ and $M_d$ can be considered OTMs that always make queries to entire prefixes of the oracle sequence, which is represented by the input string to the compression/decompression algorithm. The OTM's input $n$, which represents the size of the output prefix to compute, is then implicitly the number of bits output by $M_e$ or $M_d$. By restricting the behavior of an OTM in this way, the query usage necessarily counts all oracle bits to the left of any bit that gets queried, in addition to the queried bit. In other words, the query usage was implicitly defined by Ryabko to be the index of the rightmost queried bit, as opposed to the number of bits queried. All results of the present paper hold if query usage is instead defined in this manner.

To define a lower compression ratio, instead of considering the $\liminf_{n \to \infty}$ over all *bit* positions $n$ in $S$, which is how $\rho^-$ is defined, Ryabko considered the $\limsup_{i \to \infty}$ over all *block* positions $n_i$ (i.e. *subsequences* of bit positions), where $0 \leq n_1 < n_2 < n_3 < \ldots$. He then included the block positions as part of the specification of the $T$-code, by requiring the Turing machines to read their input and produce output in sequential blocks. Therefore the optimization over all pairs of encoding/decoding machines $M_e, M_d$ in the current paper's definition of $c_\mathrm{T}$ simultaneously optimizes over all subsequences of bit positions at which to measure the compression ratio. It is routine to verify that the infimum over all subsequences of bit positions $\{n_i\}_{i=1}^\infty$ of the $\limsup_{i \to \infty}$ over the positions $\{n_i\}_{i=1}^\infty$ is exactly the $\liminf_{n \to \infty}$ over *all* bit positions $n$.

Finally, constructive dimension as defined by Lutz [Lut03b] had not yet been defined at the time Ryabko wrote [Rya86]. He in fact showed that, for all $X \subseteq \mathbf{C}$, $c_\mathrm{T}(X) = \sup_{S \in X} \liminf_{n \to \infty} \frac{\mathrm{K}(S \restriction n)}{n}$. By Theorems 2.11 and 2.10, the right hand side is $\mathrm{cdim}(X)$.

Theorem 3.2 achieves weaker compression results than the main results of this paper, Theorems 4.3 and 4.6. Theorem 3.2 does not include $\rho^+$ or $\mathrm{cDim}$, and it requires optimizing over all OTMs. However, unlike Theorem 4.3, in which only the decompression is computable, the compression achieved in Theorem 3.2 is computable, by the definition of $c_\mathrm{T}$.

# 4 Results

We now establish the new results.

The following lemma shows two senses in which the composition of two oracle Turing machines in a transitive Turing reduction bounds the compression ratio of the transitive reduction below the product of the compression ratios of the two original reductions.

**Lemma 4.1.** *Let $S, P, R \in \mathbf{C}$ and $M_S^P, M_P^R \in \text{OTM}$ such that $S \leq_\text{T} P$ via $M_S^P$ and $P \leq_\text{T} R$ via $M_P^R$, and let $M = M_S^P \circ M_P^R$, so that $S \leq_\text{T} R$ via $M$. Then*

$$\rho_M^+(S, R) \leq \rho_{M_S^P}^+(S, P)\rho_{M_P^R}^+(P, R),$$

*and*

$$\rho_M^-(S, R) \leq \rho_{M_S^P}^-(S, P)\rho_{M_P^R}^+(P, R).$$

*Proof.* Let $r_S^{P+} > \rho_{M_S^P}^+(S, P)$, $r_S^{P-} > \rho_{M_S^P}^-(S, P)$, and $r_P^{R+} > \rho_{M_P^R}^+(P, R)$. It suffices to show that $\rho_M^+(S, R) \leq r_S^{P+}r_P^{R+}$ and $\rho_M^-(S, R) \leq r_S^{P-}r_P^{R+}$.

For infinitely many $n$, $\#_S^P(M_S^P, n) < r_S^{P-}n$. For all but finitely many $n$, $\#_S^P(M_S^P, n) < r_S^{P+}n$, and $\#_P^R(M_P^R, n) < r_P^{R+}n$. Then, for all but finitely many $n$, to compute $S \upharpoonright n$, $M$ requires

$$\#_S^R(M, n) = \#_P^R\left(M_P^R, \#_S^P\left(M_S^P, n\right)\right) < r_P^{R+}\#_S^P\left(M_S^P, n\right) < r_S^{P+}r_P^{R+}n$$

queries to $R$. Since this holds for all but finitely many $n$,

$$\rho_M^+(S, R) = \limsup_{n\to\infty} \frac{\#_S^R(M, n)}{n} \leq r_S^{P+}r_P^{R+}.$$

For infinitely many $n$, to compute $S \upharpoonright n$, $M$ requires

$$\#_S^R(M, n) = \#_P^R\left(M_P^R, \#_S^P\left(M_S^P, n\right)\right) < r_P^{R+}\#_S^P\left(M_S^P, n\right) < r_S^{P-}r_P^{R+}n$$

queries to $R$. Since this holds for infinitely many $n$,

$$\rho_M^-(S, R) = \liminf_{n\to\infty} \frac{\#_S^R(M, n)}{n} \leq r_S^{P-}r_P^{R+}.$$

$\square$

An OTM that computes a sequence $S$, together with a finite number of oracle bits that it queries, is a program to produce a prefix of $S$. Thus, the query usage of the Turing machine on that prefix cannot be far below the Kolmogorov complexity of the prefix. This is formalized in the following lemma, which bounds the compression ratio below by dimension.

**Lemma 4.2.** *Let $S, R \in \mathbf{C}$ and $M \in \text{OTM}$ such that $S \leq_\text{T} R$ via $M$. Then*

$$\rho_M^-(S, R) \geq \dim(S), \text{ and } \rho_M^+(S, R) \geq \text{Dim}(S).$$

*Proof.* Let $\pi_M$ be a self-delimiting program for $M$, so that, for all $x \in \{0,1\}^*$, $U(\pi_M x) = M(x)$. Let $r_n \in \{0,1\}^{\#_S^R(M,n)}$ be the oracle bits of $R$ queried by $M$ on input $n$, in the order in which they are queried. Recall the self-delimiting encoding function enc. For each $n \in \mathbb{N}$, let $\pi_n = \pi_{M'}\pi_M \mathrm{enc}(n)\mathrm{enc}(r_n)$, where $\pi_{M'}$ is a self-delimiting program that simulates $M$, encoded by $\pi_M$, on input $n$, encoded by $\mathrm{enc}(n)$, with oracle $R$, encoded by $\mathrm{enc}(r_n)$. When $M$ makes its $i^{\text{th}}$ query to a bit of $R$, the bit $r_n[i]$ is returned. Since $M$ queries each bit of $R$ at most once (see section 2), the bit from $r_n$ will be correct, no matter what index was queried by $M$, since the bits of $r_n$ are arranged in the order in which $M$ makes its queries.

Then $U(\pi_n) = S \upharpoonright n$, so $\mathrm{K}(S \upharpoonright n) \leq |\pi_n|$. By Theorem 2.10,

$$\begin{aligned}
\dim(S) &= \liminf_{n\to\infty} \frac{\mathrm{K}(S \upharpoonright n)}{n} \\
&\leq \liminf_{n\to\infty} \frac{|\pi_{M'}\pi_M \mathrm{enc}(n)\mathrm{enc}(r_n)|}{n} \\
&\leq \liminf_{n\to\infty} \frac{|\pi_{M'}\pi_M| + \log n + 2\log\log n + \#_R^S(M,n) + 2\log \#_R^S(M,n) + 6}{n} \\
&= \liminf_{n\to\infty} \frac{\#_R^S(M,n)}{n} \\
&= \rho_M^-(S,R),
\end{aligned}$$

and similarly, $\mathrm{Dim}(S) \leq \rho_M^+(S,R)$. $\square$

The next theorem is the main result of this paper. It shows that the compression lower bounds of Lemma 4.2 are achievable, and that a single OTM $M$ suffices to carry out the reduction, no matter which sequence $S$ is being computed. Furthermore, the oracle sequence $R$ to which $S$ reduces can be made Martin-Löf random. The randomness of $R$ is easily accomplished by invoking the construction of Gács in a black-box fashion; the majority of the work in the proof is establishing the bound on the compression.

**Theorem 4.3.** *There is an OTM $M$ such that, for all $S \in \mathbf{C}$, there is a sequence $R \in$ RAND such that*

1. $S \leq_\mathrm{T} R$ via $M$.

2. $\rho_M^-(S,R) = \dim(S)$.

3. $\rho_M^+(S,R) = \mathrm{Dim}(S)$.

**Proof idea:** If the dimension of $S$ is small, then the optimal constructive martingale $\mathbf{d}$ performs well on $S$. Thus, if we have already computed a prefix $S \upharpoonright n$ of $S$, then *on average*, $\mathbf{d}$ increases its capital more on the next $k$ bits of $S$ than it would on other $k$-bit strings that could extend $S \upharpoonright n$. This places the next $k$ bits of $S$ in a small (on average) subset of $\{0,1\}^k$, namely, those strings on which $\mathbf{d}$ increases its capital above a certain threshold $d_n$, which is chosen to be slightly smaller than $\mathbf{d}(S \upharpoonright (n+k))$, the amount of capital made after the next $k$ bits of $S$. Since $\mathbf{d}$ is constructive, it is possible to enumerate strings from this small set by evaluating the computable function $\widehat{\mathbf{d}}$ in parallel on all possible length-$k$ extensions of $S \upharpoonright n$, and outputting a string $u \in \{0,1\}^k$ when $\widehat{\mathbf{d}}((S \upharpoonright n)u, t)$ is greater than $d_n$, for some value of $t \in \mathbb{N}$. We will encode the next $k$ bits of $S$ as an index into this set, where the index will represent the order in which this parallel evaluation enumerates the string we want – the next $k$ bits of $S$. This technique is similar to that used by Merkle and Mihailović [MM04] to prove Theorem 3.1.

We require two lemmas to prove Theorem 4.3. Lemma 4.4 shows that the average number of bits needed to encode the index of a length-$k$ extension of $S \upharpoonright n$ is close to the dimension of $S$ times $k$. We will also need to encode the threshold $d_n$ into the oracle sequence, since the actual amount of capital that $\mathbf{d}$ will make is uncomputable. Lemma 4.5 shows that we can find a rational threshold $d_n$ that requires so few bits to represent that it will not affect the compression ratio when added to the oracle sequence, yet which is still a close enough approximation to $\mathbf{d}(S \upharpoonright (n+k))$ to keep the index length of Lemma 4.4 small.

**Lemma 4.4.** *Let $S \in \mathbf{C}$. For all $i \in \mathbb{N}$, define $k_i = i+1$, and define $n_0 = 0$ and $n_i = n_{i-1} + k_i = \frac{i(i+1)}{2}$ for $i > 0$. Let $d_0, d_1, \ldots$ be a sequence of real numbers such that, for all $i \in \mathbb{N}$, $d_i \geq \mathbf{d}(S \upharpoonright n_i)\left(1 - \frac{1}{i^2}\right)$. Define $A_i \subseteq \{0,1\}^{k_i}$ by*

$$A_i = \left\{ u \in \{0,1\}^{k_i} \mid \mathbf{d}((S \upharpoonright n_{i-1})u) > d_i \right\}.$$

*Then*

$$\liminf_{i \to \infty} \frac{\sum_{j=0}^{i} \log |A_j|}{n_i} \leq \dim(S), \text{ and } \limsup_{i \to \infty} \frac{\sum_{j=0}^{i} \log |A_j|}{n_i} \leq \mathrm{Dim}(S).$$

*Proof.* We show the result for $\dim(S)$. The proof for $\mathrm{Dim}(S)$ is similar, replacing "for infinitely many $i$" conditions with "for all but finitely many $i$."

The indices $n_0 < n_1 < n_2 < \ldots$ partition $S$ into blocks $S[n_0 \ldots n_1 - 1]$, $S[n_1 \ldots n_2 - 1]$, $\ldots$, with $k_i = n_{i+1} - n_i$ equal to the length of the $i^{\text{th}}$ block, and $n_i$ equal to the length of the first $i+1$ blocks.

Let $t' > t > \dim(S)$. It suffices to show that, for infinitely many $i \in \mathbb{N}$, $\sum_{j=0}^{i} \log |A_j| \leq t' n_i$. Since $t > \dim(S)$, for infinitely many $n \in \mathbb{N}$,

$$\mathbf{d}(S \upharpoonright n) \geq 2^{(1-t)n} \mathbf{d}(\lambda).$$

A martingale can at most double its capital after every bit, and each index $n$ with $n_i \leq n < n_{i+1}$ is at most $k_i$ bits beyond $n_i$. It follows that for infinitely many $i \in \mathbb{N}$,

$$\mathbf{d}(S \upharpoonright n_i) \geq 2^{(1-t)n_i - k_i} \mathbf{d}(\lambda). \tag{4.1}$$

For all $i \in \mathbb{N}$, set $l_i \in \mathbb{R}$ such that $\mathbf{d}(S \upharpoonright n_i) = 2^{k_i - l_i} \mathbf{d}(S \upharpoonright n_{i-1})$. By induction on $i$,

$$\mathbf{d}(S \upharpoonright n_i) = \mathbf{d}(\lambda) \prod_{j=0}^{i} 2^{k_j - l_j}. \tag{4.2}$$

Then, by equations (4.1) and (4.2), and the fact that $\sum_{j=0}^{i-1} k_i = n_i$, for infinitely many $i \in \mathbb{N}$,

$$\prod_{j=0}^{i} 2^{k_j - l_j} \geq 2^{(1-t)n_i - k_i} \implies \sum_{j=0}^{i}(k_j - l_j) \geq (1-t)n_i - k_i \implies \sum_{j=0}^{i} l_j \leq tn_i + 2k_i.$$

Recall that $\mathbf{d}(S \restriction n_i)\left(1 - \frac{1}{i^2}\right) \leq d_i$. By Corollary 2.5 (take $k = k_i, l = l_i, \alpha = 1 - \frac{1}{i^2}, w = S \restriction n_{i-1}$) and the definition of $l_i$, since

$$d_i \geq \left(1 - \frac{1}{i^2}\right) \mathbf{d}(S \restriction n_i) = \left(1 - \frac{1}{i^2}\right) 2^{k_i - l_i} \mathbf{d}(S \restriction n_{i-1}),$$

it follows that $|A_i| \leq \frac{2^{l_i}}{1 - \frac{1}{i^2}}$, and so $\log |A_i| \leq l_i - \log\left(1 - \frac{1}{i^2}\right)$. Let $c_{0,1} = \log |A_0| + \log |A_1| - l_0 - l_1$. Then

$$\sum_{j=0}^{i} \log |A_j| \leq \sum_{j=0}^{i} l_j - \sum_{j=2}^{i} \log\left(1 - \frac{1}{j^2}\right) + c_{0,1}$$

$$\leq tn_i + 2k_i - \sum_{j=2}^{i} \underbrace{(\log(j+1) + \log(j-1) - 2\log j)}_{\text{telescopes}} + c_{0,1}$$

$$= t'n_i + (t - t')n_i + 2k_i - (\log 1 - \log 2 - \log i + \log(i+1)) + c_{0,1}.$$

$t < t'$, $2k_i = o(n_i)$, and $\lim_{i \to \infty}[\log(i+1) - \log i] = 0$. Therefore, for infinitely many $i$, $\sum_{j=0}^{i} \log |A_j| \leq t'n_i$. □

**Lemma 4.5.** *Let $i \in \mathbb{Z}^+$ $c \in \mathbb{R}^+$, and $r \in \left[1, c2^{i^2}\right]$. Then there is a rational number $d \in \mathbb{Q}^+$ such that $r > d \geq r\left(1 - \frac{1}{i^2}\right)$ and $\mathrm{K}(d) = O(\log i)$.*

*Proof.* We prove the cases $r \geq i^2$ and $1 \leq r < i^2$ separately. Suppose $r \geq i^2$. In this case we will choose $d$ to be an integer. Set $k \in \mathbb{Z}^+$ such that $2^{k-1} < i^2 \leq 2^k$. Since $r \geq i^2 > 2^{k-1}$, $\lceil \log r \rceil > k - 1$.

Let $d \in \mathbb{Z}^+$ be the integer whose binary representation is $x0^{\lceil \log r \rceil - k}$, where $x \in \{0,1\}^k$ is the first $k$ bits of $\lfloor r \rfloor$. Since $d$ shares its first $k$ bits with $r$,

$$r - d \leq 2^{\lceil \log r \rceil - k} - 1 \leq \frac{r+2}{2^k} - 1 \leq \frac{r}{i^2},$$

so $r > d \geq r\left(1 - \frac{1}{i^2}\right)$. $d$ can be fully described by the first $k$ bits of $r$, along with the binary representation of the number $\lceil \log r \rceil - k$ of 0's that follow. Thus, describing $d$ requires no more than $k + \log(\lceil \log r \rceil - k) \leq \log i^2 + 1 + \log \log c + \log i^2 = O(\log i)$ bits.

This will not work if $r \in \mathbb{Z}^+$ and $r$'s least significant $\lceil \log r \rceil - k$ bits are 0, which would result in $d = r$, rather than $d < r$. In this case, let

$$d = r - 1 = \mathrm{bnum}\left(\mathrm{rep}_2(\mathrm{bnum}(x) - 1)1^{\lceil \log r \rceil - k}\right),$$

where $\text{bnum}(x)$ is the integer whose binary representation is $x$, and $\text{rep}_2(n)$ is the binary representation (with possible leading zeroes) of $n \in \mathbb{N}$. This likewise requires $O(\log i)$ bits to describe. Since $r \geq i^2$, $d = r - 1 \geq r\left(1 - \frac{1}{i^2}\right)$.

Now suppose that $1 \leq r < i^2$. We approximate $r$ by the binary integer $\lfloor r \rfloor$, plus a finite prefix of the bits to the right of $r$'s decimal point in binary form. If $x.S$ is the binary representation of $r$, where $x \in \{0,1\}^*$ and $S \in \mathbf{C}$, let $d \in \mathbb{Z}^+$ be represented by $x.y$, where $y \sqsubseteq S$.

Since $r < i^2$, $|x| \leq \log i^2 = O(\log i)$. We need $r - d \leq \frac{r}{i^2}$ for $d$ to approximate $r$ closely. Since $r - d \leq 2^{-|y|}$, it suffices to choose $y \sqsubseteq S$ such that $2^{-|y|} \leq \frac{r}{i^2}$, or $|y| \geq \log \frac{i^2}{r}$. Let $|y| = \left\lceil \log \frac{i^2}{r} \right\rceil = O(\log i)$, since $r \geq 1$. Thus $|x| + |y| = O(\log i)$, so describing $d$ requires $O(\log i)$ bits.

This will not work if $r$ is a dyadic rational $x.z$, where $x, z \in \{0,1\}^*$ and $|z| \leq |y|$, which would result in $d = r$, rather than $d < r$. In this case, let $r' \in \left[r\left(1 - \frac{1}{2i^2}\right), r\right)$ be irrational. Choose $d$ for $r'$ by the method just described, such that $r' > d \geq r'\left(1 - \frac{1}{2i^2}\right)$, and $d$ requires $O(\log(i\sqrt{2})) = O(\log i)$ bits. Then $d \geq r\left(1 - \frac{1}{i^2}\right)$ by the triangle inequality, and $d < r' < r$. $\square$

Finally, we prove Theorem 4.3.

*Proof of Theorem 4.3.* If $S \in \text{RAND}$, then $S \leq_{\text{T}} S$ via the trivial "bit copier" machine $M'$, with lower and upper compression ratio $\dim(S) = \text{Dim}(S) = 1$, so assume that $S \notin \text{RAND}$.

A single OTM $M''$ suffices to carry out the reduction described below, no matter what sequence $S \notin \text{RAND}$ is being computed. If $S \in \text{RAND}$, then $M'$ is used. These two separate reductions are easily combined into one by reducing each sequence $S$ to a random sequence $bR$ via $M \in \text{OTM}$, where $b \in \{0,1\}$, $R = S$ if $S \in \text{RAND}$, and $R$ is given by the construction below if $S \notin \text{RAND}$. The bit $b$ indicates to $M$ whether to use $M'$ or $M''$ for the reduction. Hence a single OTM $M$ implements the "optimal decompression".

For all $i \in \mathbb{N}$, define $k_i = i + 1$, and define $n_0 = 0$ and $n_i = n_{i-1} + k_i = \frac{i(i+1)}{2}$ for $i > 0$. Note that $n_i \leq i^2$ for all $i \geq 3$. $k_i$ represents the length of the $i^{\text{th}}$ block into which we subdivide $S$. $n_i$ is the total length of the first $i+1$ blocks. Define $d_i \in \mathbb{Q}^+$ to be a rational number satisfying

1. $\mathbf{d}(S \upharpoonright n_i)\left(1 - \frac{1}{i^2}\right) \leq d_i < \mathbf{d}(S \upharpoonright n_i)$; i.e., $d_i$ is a rational number approximating $\mathbf{d}(S \upharpoonright n_i)$ from below.

2. $\text{K}(d_i) = o(k_i)$; i.e. $d_i$ can be computed from a program asymptotically smaller than the length of the $i^{\text{th}}$ block.

By Observation 2.3, $\mathbf{d}(S \upharpoonright n_i) \leq 2^{n_i}\mathbf{d}(\lambda) \leq 2^{i^2}\mathbf{d}(\lambda)$ for $i \geq 3$. By Theorem 2.6, $S \notin \text{RAND}$ implies that for all but finitely many $i$, $\mathbf{d}(S \upharpoonright n_i) \geq 1$. Thus, by Lemma 4.5 (take $r = \mathbf{d}(S \upharpoonright n_i)$ and $c = \mathbf{d}(\lambda)$), there is a $d_i \in \mathbb{Q}^+$ satisfying the above two conditions.

Define the set $A_i \subseteq \{0,1\}^{k_i}$ for all $i \in \mathbb{N}$ as in Lemma 4.4 by

$$A_i = \left\{ u \in \{0,1\}^{k_i} \mid \mathbf{d}((S \upharpoonright n_{i-1})u) > d_i \right\},$$

the set of all length-$k_i$ extensions of $S \upharpoonright n_{i-1}$ that add more capital to the optimal constructive martingale $\mathbf{d}$ than $S[n_{i-1} \ldots n_i - 1]$ does, to within multiplicative factor at most $1 - \frac{1}{i^2}$. Since $\mathbf{d}(S \upharpoonright n_i) > d_i$, it follows that $S[n_{i-1} \ldots n_i - 1] \in A_i$.

For all $i \in \mathbb{N}$, let $p_i \in \mathbb{N}$ be the output of the following partial computable procedure, when given as input the string $S[n_{i-1} \ldots n_i - 1] \in \{0,1\}^{k_i}$:

STRING-TO-INDEX($S[n_{i-1} \ldots n_i - 1]$)

1  $A_i \leftarrow \varnothing$
2  **for** $t = 0, 1, 2, \ldots$
3     **do for** each $u \in \{0,1\}^{k_i} - A_i$
4        **do if** $\widehat{\mathbf{d}}((S \upharpoonright n_{i-1})u, t) > d_i$
5           **then** add $u$ to $A_i$
6              **if** $u = S[n_{i-1} \ldots n_i - 1]$
7                 **then** output $|A_i|$ and halt

In other words, $p_i$ is the order in which $\mathbf{d}(S \upharpoonright n_i)$ is shown to exceed $d_i$ (i.e., to belong to $A_i$) by a parallel evaluation of $\widehat{\mathbf{d}}((S \upharpoonright n_{i-1})u, t)$ on all extensions $u \in \{0,1\}^{k_i}$ of $S \upharpoonright n_{i-1}$, for $t = 0, 1, 2, \ldots$. Since $d_i < \mathbf{d}(S \upharpoonright n_i)$, there exists some $t \in \mathbb{N}$ such that $\widehat{\mathbf{d}}(S \upharpoonright n_i, t) > d_i$, and so $p_i$ is well-defined. The computation of INDEX-TO-STRING, the inverse of STRING-TO-INDEX, resembles that of STRING-TO-INDEX:

INDEX-TO-STRING($p_i$)

1  $A_i \leftarrow \varnothing$
2  **for** $t = 0, 1, 2, \ldots$
3     **do for** each $u \in \{0,1\}^{k_i} - A_i$
4        **do if** $\widehat{\mathbf{d}}((S \upharpoonright n_{i-1})u, t) > d_i$
5           **then** add $u$ to $A_i$
6              **if** $|A_i| = p_i$
7                 **then** output $u$ and halt

Note that INDEX-TO-STRING will not halt if given as input an integer greater than $|A_i|$, and STRING-TO-INDEX will not halt if given a string that is not an element of $A_i$.

For all $i \in \mathbb{N}$, let $\pi(d_i)$ denote a self-delimiting, shortest program for computing $d_i$. Define the sequence $P \in \mathbf{C}$ by

$$P = \text{enc}(p_0)\pi(d_0)\text{enc}(p_1)\pi(d_1)\text{enc}(p_2)\pi(d_2)\ldots.$$

Define the oracle Turing machine $M_S^P$ that produces $n$ bits of $S$, with oracle $P$, as follows. Let $i(n)$ denote the block in which $n$ resides – the unique $i \in \mathbb{N}$ such that $n_i \leq n < n_{i+1}$. First, $M_S^P$ reads the first $i(n) + 1$ blocks of $P$:

$$\text{enc}(p_0)\pi(d_0)\ldots\text{enc}(p_{i(n)})\pi(d_{i(n)}).$$

$M_S^P$ then calculates the first $i(n)+1$ blocks of $S$ iteratively. On block $i$, $M_S^P$ first computes $p_i$ from $\text{enc}(p_i)$ and $d_i$ from $\pi(d_i)$. Then, $M_S^P$ evaluates INDEX-TO-STRING($p_i$) to obtain $S[n_{i-1} \ldots n_i]$ and outputs it as the $i^{\text{th}}$ block of $S$.

Since $S[n_{i-1}..n_i - 1] \in A_i$, it follows that $p_i \leq |A_i|$, and so $|\text{enc}(p_i)| \leq \log|A_i| + 2\log\log|A_i| + 3$. Therefore, by Lemma 4.4,

$$\liminf_{i \to \infty} \frac{\sum_{j=0}^{i} |\text{enc}(p_j)|}{n_i} \leq \liminf_{i \to \infty} \frac{\sum_{j=0}^{i} \log|A_j|}{n_i} \leq \dim(S), \quad (4.3)$$

and

$$\limsup_{i \to \infty} \frac{\sum_{j=0}^{i} |\text{enc}(p_j)|}{n_i} \leq \limsup_{i \to \infty} \frac{\sum_{j=0}^{i} \log|A_j|}{n_i} \leq \text{Dim}(S). \quad (4.4)$$

By our choice of $d_i$, $|\pi(d_i)| = o(k_i)$, so $\sum_{j=0}^{i} |\pi(d_j)| = o(n_i)$, giving

$$\liminf_{i \to \infty} \frac{\sum_{j=0}^{i} |\text{enc}(p_j)\pi(d_j)|}{n_i} = \liminf_{i \to \infty} \frac{\sum_{j=0}^{i} |\text{enc}(p_j)|}{n_i}, \quad (4.5)$$

and

$$\limsup_{i \to \infty} \frac{\sum_{j=0}^{i} |\text{enc}(p_j)\pi(d_j)|}{n_i} = \limsup_{i \to \infty} \frac{\sum_{j=0}^{i} |\text{enc}(p_j)|}{n_i}. \quad (4.6)$$

By the definition of lim inf,

$$\liminf_{n \to \infty} \frac{\sum_{j=0}^{i(n)} |\text{enc}(p_j)\pi(d_j)|}{n} \leq \liminf_{i \to \infty} \frac{\sum_{j=0}^{i} |\text{enc}(p_j)\pi(d_j)|}{n_i}. \quad (4.7)$$

Since $n_i = \frac{k_i(k_i+1)}{2}$, $k_i = o(n_i)$, so

$$\limsup_{n \to \infty} \frac{\sum_{j=0}^{i(n)} |\text{enc}(p_j)\pi(d_j)|}{n} \leq \limsup_{i \to \infty} \frac{\sum_{j=0}^{i} |\text{enc}(p_j)\pi(d_j)|}{n_i}. \quad (4.8)$$

In other words, because the block size grows slower than the prefix length, the lim sup over all blocks is at least the lim sup over all bits (and they are in fact equal by the definition of lim sup). Regardless of the block growth rate, this inequality holds trivially for lim inf.

For all $n \in \mathbb{N}$, $M_S^P$ requires $\sum_{j=0}^{i(n)} |\text{enc}(p_j)\pi(d_j)|$ bits of $P$ in order to compute $n$ bits of $S$, and hence, by inequalities (4.3)-(4.8),

$$\rho_{M_S^P}^{-}(S, P) = \liminf_{n \to \infty} \frac{\sum_{j=0}^{i(n)} |\text{enc}(p_j)\pi(d_j)|}{n} \leq \dim(S),$$

and

$$\rho_{M_S^P}^{+}(S, P) = \limsup_{n \to \infty} \frac{\sum_{j=0}^{i(n)} |\text{enc}(p_j)\pi(d_j)|}{n} \leq \text{Dim}(S).$$

Let $R \in \text{RAND}$ and $M_P^R \in \text{OTM}$ be given by the construction of Gács in his proof of Theorem 3.1, satisfying $P \leq_\text{T} R$ via $M_P^R$ and $\rho_{M_P^R}^{+}(P, R) = 1$. Let $M'' = M_S^P \circ M_P^R$. Then $S \leq_\text{T} R$ via $M''$ and, by Lemma 4.1,

$$\rho_{M''}^{-}(S, R) \leq \rho_{M_S^P}^{-}(S, P)\rho_{M_P^R}^{+}(P, R) \leq \dim(S),$$

and
$$\rho^+_{M''}(S, R) \leq \rho^+_{M^P_S}(S, P)\rho^+_{M^R_P}(P, R) \leq \text{Dim}(S).$$

By Lemma 4.2, $\rho^-_{M''}(S, R) \geq \dim(S)$ and $\rho^+_{M''}(S, R) \geq \text{Dim}(S)$. $\square$

Finally, these results give a new characterization of constructive dimension.

**Theorem 4.6.** *For every sequence $S \in \mathbf{C}$,*

$$\dim(S) = \rho^-(S), \text{ and } \text{Dim}(S) = \rho^+(S),$$

*and, for all $X \subseteq \mathbf{C}$,*

$$\text{cdim}(X) = \sup_{S \in X} \rho^-(S), \text{ and } \text{cDim}(X) = \sup_{S \in X} \rho^+(S).$$

*Proof.* Immediate from Lemma 4.2 and Theorems 4.3 and 2.11. $\square$

It is instructive to compare Theorem 4.6 with Ryabko's Theorem 3.2, considering especially what they say about individual sequences. While Ryabko's theorem represents $S$ with a more compact sequence $R$, it is not optimally compact, as a different decoding machine is required to get the compression ratio closer and closer to the optimal ratio of $\dim(S)$. However, the major difference between the theorems is that Ryabko's construction does not achieve the bound between $\rho^+$ and Dim. Intuitively, Ryabko's theorem states that $S$ may be compressed to a sequence $R$, where *infinitely often* (but not almost everywhere), approximately the first $K(S \upharpoonright n)$ bits of $R$ suffice to produce $S \upharpoonright n$. However, Ryabko's construction requires that the block lengths grow exponentially, so that if $S$ is written $x_1 x_2 x_3 \ldots$, then for all $i \in \mathbb{N}$, $|x_1 \ldots x_i| < 2^{-i}|x_1 \ldots x_{i+1}|$. Therefore, while the lower compression ratio $\rho^-$ is close to optimal, the upper compression ratio $\rho^+$ goes to infinity.

# 5 Conclusion

We have shown that every infinite sequence is Turing reducible to a Martin-Löf random infinite sequence with the optimal compression ratio possible. Since this optimal ratio is the constructive dimension of the sequence, this gives a new characterization of constructive dimension in terms of Turing reduction compression ratios.

The Turing reductions of Theorems 3.1, 3.2, and 4.3 satisfy the stronger properties of the *weak truth-table reduction* (see [Soa87]), which is a Turing reduction in which the query usage of the reduction machine $M$ on input $n$ is bounded by a computable function of $n$. For example, $1.01n + O(1)$ suffices. Thus, constructive dimension could also be defined in terms of weak truth-table reductions.

As noted in the introduction, for the sequences $S$ and $R$ in Theorems 3.1 and 4.3, it is not necessarily the case that $R \leq_T S$. In other words, though the decompression is computable, it is not computably reversible in all cases. For instance, if $S$ is computable, then $R \not\leq_T S$, since no sequence $R \in \text{RAND}$ is computable. For this reason, Theorem 4.3 does not imply Theorem 3.2, which allows for the reduction to be computably reversed, subject to the trade-off that the compression requirements are weakened. It remains open whether the compression direction is computable if we drop the requirement that the sequence $R$ be random.

**Acknowledgment.** I am grateful to Philippe Moser and Xiaoyang Gu for their insightful discussions, to Jack Lutz and Jim Lathrop for their helpful advice in preparing this article, and to John Hitchcock for making useful corrections in an earlier draft. I also thank anonymous referees for helpful suggestions.